\documentclass[a4paper,11pt]{article}
\pdfoutput=1 % if your are submitting a pdflatex (i.e. if you have
\usepackage{jheppub}
\usepackage{epsfig}
\usepackage{amsmath}
\usepackage{graphicx}
\usepackage{multirow}
\usepackage{cancel}
\usepackage{extarrows}
\def\A0#1{\Pi_{\rm #1}(0)}
\def\AP0#1{\Pi'_{\rm #1}(0)}

\def\be{\begin{equation}}
\def\ee{\end{equation}}
\def\bea{\begin{array}}
\def\eea{\end{array}}
\def\beqa{\begin{eqnarray}}
\def\eeqa{\end{eqnarray}}
\def\beqas{\begin{eqnarray*}}
\def\eeqas{\end{eqnarray*}}

\def\bp{\begin{picture}}
\def\ep{\end{picture}}
\def\bc{\begin{center}}
\def\ec{\end{center}}
\def\bfig{\begin{figure}}
\def\efig{\end{figure}}

\def\bit{\begin{itemize}}
\def\eit{\end{itemize}}
\def\nn{\nonumber}
\def\f{\frac}

\def\[{\left[}
\def\]{\right]}
\def\({\left(}
\def\){\right)}

\def\..{\left.}
\def\.{\right.}
\def\tl{\tilde}
\def\ra{\rightarrow}
\def\la{\leftarrow}

\def\tm{\times}

\def\da{\dagger}

\def\la{\lambda}

\def\al{\alpha}
\def\bt{\beta}

\def\ep{\epsilon}

\def\Ga{\Gamma}
\def\ga{\gamma}

\def\pr{\prime}

\title{ Flipped SU(6) Unification of the Sequential $SU(3)_c\tm SU(3)_L\tm U(1)_X$ Model}
\author[a]{Fei Wang}
\affiliation[a]{School of Physics, Zhengzhou University, Zhengzhou 450000, P. R. China}
\emailAdd{feiwang@zzu.edu.cn}
\abstract{We propose to partially unify the sequential $SU(3)_c\tm SU(3)_L\tm U(1)_X$ model (with $\beta=1/\sqrt{3}$) into the flipped $SU(6)$ model with the gauge group $SU(6)\tm U(1)_K$. Gauge anomaly cancellation can easily be satisfied. We discuss the relevant Higgs sector, the low energy $\emph{331}$ model spectrum  and the unification of $SU(3)_c$ and $SU(3)_L$ gauge couplings. Neutrino masses generation and successful gauge coupling unification can set lower/upper bounds on the $\emph{331}$ breaking scale. The partial proton decay lifetime of various channels, for example, the $p\to e^+ \pi^0$ channel, in flipped SU(6) GUT are discussed.  We find that certain parameter region with $M_{331}\sim 10^{15}$ GeV of case II (for case with $M_{331}$ scale ${\bf \tl{H}_{3,8}}$ Higgs field) can predict a partial proton lifetime of order $10^{34}$ years for $p\to e^+ \pi^0$ mode, which can be tested soon by future DUNE and Hyper-Kamiokande experiments. }
\begin{document}
\maketitle
\indent
\newpage
\section{Introduction}

 The standard model (SM) of particle physics, based on the $SU(3)_c\tm SU(2)_L\tm U(1)_Y$ gauge group, has been extremely successful in describing phenomena below the weak scale. However, the SM still leaves some theoretical and aesthetical questions unanswered, for example, the origin of charge quantization, the values of the low energy parameters and the origin of the flavor structures. Such questions can be answered in the framework of Grand Unified Theory (GUT), such as $SU(5)$~\cite{su5} and $SO(10)$~\cite{so10} GUT. In the GUT framework, the matter contents of SM can be embedded into certain representations of the GUT group, indicating that the low energy Yukawa couplings can be obtained from a single Yukawa coupling (or few Yukawa couplings) at the GUT scale. The approximate unification of the SM couplings strongly indicate the existence of GUT. %%The low energy gauge couplings can be the consequence of renormalization group equation (RGE) evolution a single GUT group gauge couplings from the unification scale to M_Z.
 We know that the SU(5) GUT model unified the SM gauge group directly at the GUT scale without any intermediate partial unification step. If intermediate partial unification exists at a higher scale beyond $M_Z$, for example, the Pati-Salam $SU(4)_c\tm SU(2)_L\tm SU(2)_R$ partial unification, genuine gauge couplings unification needs a larger GUT group, such as $SO(10)$ in this case. So, it is interesting to seek other unification model with some intermediate partial unification steps, such as the (partial) GUT model with an intermediate $SU(3)_c\tm SU(3)_L\tm U(1)_X$ partial unification~\cite{331model:1,331model:2,331model:A,331model:B,331model:C} step.

The measured value of the electroweak mixing angle $\sin^2\theta_W(M_Z)=0.23\lesssim0.25$ appears to obey an $SU(3)$ symmetry in such a way that $\sin^2\theta_W(\mu)= 1/4$ at some new fundamental energy scale $\mu$ upon TeV~\cite{weinberg}. By introducing an extra $U(1)$ factor to accommodate quark sector, one can arrive at an $SU(3)_c\tm SU(3)_L\tm U(1)_X$ model ($\emph{331}$ model). Depending on different choices of the $\beta$ value ($\beta=1/\sqrt{3}$~\cite{331model:X,331model:3} or $\beta=\sqrt{3}$~\cite{331model:X,331model:4,331model:5}) within the embedding of the electric charge, $\emph{331}$ models in the literatures need to introduce different electrically charged particles for the fitting of the $SU(3)_L$ representations. It is remarkable that the existence of three matter generations could be the consequence of gauge anomaly cancellation requirements. Besides, the heaviness of the top quark mass and the emergence of the Peccei-Quinn (PQ) symmetry can also possibly be explained in the $\emph{331}$ framework~\cite{331:PQ1,331:PQ2}.

To understand the origin of charge quantization and the values of the low energy parameters, the intermediate $\emph{331}$ model needs to be unified into a true GUT theory. The unification of sequential $\emph{331}$ model into SU(6) model had been proposed in~\cite{su6GUT1,su6GUT2} and studied in~\cite{su6GUT3,su6GUT4,su6GUT5}. On the other hand, the genuine unification of $\emph{331}$ into $SU(6)$ needs the introduction of additional adjoint fermions and scalars etc at some intermediate scale between the $331$ scale (at ${\cal O}$ (TeV)) and GUT scale~\cite{su6GUT1}, reducing the predicability of the GUT theory. So, it is interesting to seek alternative (partial) unification steps for the $\emph{331}$ models.

We propose to partially unify the $\emph{331}$ gauge group into flipped $SU(6)$ with the gauge group $SU(6)\tm U(1)_K$.
Similar to the flipped $SU(5)$ model~\cite{fSU5:1,fSU5:2,fSU5:3}, the flipped $SU(6)$ model can be  well motivated from string theory models~\cite{fSU5:A,fSU5:B,Jiang:2009za}, which uses level-one Kac-Moody algebras and do not need adjoint Higgs fields for symmetry breaking~\cite{fSU5:4}. We know that the flipped SU(5) can adopt the economical missing-partner mechanism and possibly provide an unified cosmological scenario for inflation, dark matter and baryogenesis etc~\cite{fSU5:5,fSU5:6,fSU5:7}. We anticipate that such virtues can also be present for flipped $SU(6)$. Flipped SU(6) model can also be unified into SO(12) or $E_6$ GUT via intermediate $SU(6)\tm SU(2)$ step.

This paper is organized as follows. In Sec~\ref{sec-2}, we brief review the sequential $\emph{331}$ model and discuss the embedding of such $\emph{331}$ model into flipped SU(6) GUT model. In Sec~\ref{sec-3}, we discuss various sub-scenarios of $\emph{331}$ model and the corresponding gauge coupling unification. In Sec~\ref{sec-4}, we discuss the triggered proton decay modes and lifetimes in flipped SU(6). Sec \ref{sec-5} contains our conclusions.

\section{\label{sec-2}$SU(3)_c\tm SU(3)_L\tm U(1)_X$ unification into $SU(6)\tm U(1)_K$ model}

\subsection{Brief review of the $\emph{331}$ model with $\beta=1/\sqrt{3}$}
Ordinary (non-sequential) $\emph{331}$ model with $\beta=1/\sqrt{3}$ assigns different $SU(3)_c\tm SU(3)_L\tm U(1)_X$ quantum numbers for the three generations. The filling of the  matter contents are given as
\beqa
&& Q_{iL}{\bf (3, 3, 0)}\sim
\left(
\begin{array}{c}
U_L\\
D_L\\
(XD)_L
\end{array}\right),
~~Q_{3L}{\bf (3, \bar{3}, \f{1}{3})}\sim
\left(
\begin{array}{c}
b_L\\
t_L\\
(XT)_L
\end{array}\right),\nn\\
&& F_{aL}{\bf (1, \bar{3}, -\f{1}{3})} \sim \left(
\begin{array}{c}
E_L\\
-\nu_L\\
N_L^s
\end{array}\right),~~E_{aL}^c\sim {\bf (1, 1, 1)}~,
\eeqa
 and
\beqa
U_{iL}^c \sim {\bf (\bar{3}, 1, -\f{2}{3})}, \
D_{iL}^c \sim {\bf (\bar{3}, 1, \f{1}{3})},  \  (XD)_{iL}^c \sim {\bf (\bar{3}, 1, \f{1}{3})},\nn\\
t_{iL}^c \sim {\bf (\bar{3}, 1, -\f{2}{3})}, \
b_{iL}^c \sim {\bf (\bar{3}, 1, \f{1}{3})},  \  (XT)_{L}^c \sim {\bf (\bar{3}, 1, -\f{2}{3})},
\eeqa
 with $a=1,2,3$ and $i=1,2$ the family indices. Here $(XD)_{iL},(XD)_{iL}^c$ and $(XT), (XT)_{L}^c$  denotes some exotic vector-like quarks with the SM quantum numbers
\beqa
&&(XD)_{iL}: {\bf (3,1,-\f{1}{3})},~~~~~~~(XT)_{L}: {\bf (3,1,\f{2}{3})},\nn\\
&&(XD)_{iL}^c: {\bf (\bar{3},1,\f{1}{3})},~~~~~~~~(XT)_{L}^c: {\bf (\bar{3},1,-\f{2}{3})}.
\eeqa
 We adopt the notation $N_L^c\equiv (N^c)_L\equiv (N_R)^c$, where $\psi^c=C\bar{\psi}^T$ and $C$ the charge charge conjugate matrix. The relation of the hypercharge to the $SU(3)_L\tm U(1)_X$ generators is given by
\beqa
Y=\f{1}{\sqrt{3}}T_8 + X~,
\eeqa
with the choice of $T_8$ for fundamental representation ${\bf 3}$ of $SU(3)_L$ as
\beqa
T_8=\f{1}{2\sqrt{3}}{\rm diag}(-1,-1,~2).
\eeqa
It can be checked that the gauge anomaly can cancel only if we take into account the contributions from all the generations.

Ordinary $\emph{331}$ model contains a simple lepton sector and can be potentially tested in the TeV scale. There is an interesting variant $\emph{331}$ model called sequential $\emph{331}$ model~\cite{su6GUT1}, which, unlike ordinary $\emph{331}$ models, assign identically the matter quantum numbers for the three generations.  Therefore, the gauge anomalies are canceled for each generation separately. The filling of the  matter contents in sequential $SU(3)_c\tm SU(3)_L\tm U(1)_X$ model with $\beta=1/\sqrt{3}$ is given by
\beqa
F_L{\bf (1, \bar{3}, -\f{1}{3})} \sim \left(
\begin{array}{c}
E_L\\
-\nu_L\\
N_L^s
\end{array}\right),
~~~
Q_L{\bf (3, 3, 0)}\sim
\left(
\begin{array}{c}
U_L\\
D_L\\
(XD)_L
\end{array}\right),\nn\\
\tl{X}_L{\bf (1, \bar{3}, -\f{1}{3})}\sim
\left(
\begin{array}{c}
(XE)_L\\
(XN)_L\\
N_L
\end{array}\right),~~~\tl{Y}_L{\bf (1, \bar{3}, \f{2}{3})}\sim
\left(
\begin{array}{c}
(XN)^c_L\\
(XE)_L\\
E_L^c
\end{array}\right),
\eeqa
and
\beqa
U_{L}^c \sim {\bf (\bar{3}, 1, -\f{2}{3})}, \
D_{L}^c \sim {\bf (\bar{3}, 1, \f{1}{3})},  \  (XD)_{L}^c \sim {\bf (\bar{3}, 1, \f{1}{3})}.
\eeqa
with the SM quantum numbers for some exotic vector-like quarks and leptons $(XL)_L$ and $(XL)_L^c$ given by
\beqa
&&(XL)_L: {\bf (1,\bar{2},-\f{1}{2})},~~~~~~(XD)_L: {\bf (3,1,\f{1}{3})}~,\nn\\
&&(XL)_L^c: {\bf (1,\bar{2},~\f{1}{2})},~~~~~~~(XD)_L^c: {\bf (\bar{3},1,-\f{1}{3})}~.
\eeqa
The relation of the hypercharge to the $SU(3)_L\tm U(1)_X$ generators is the same as the non-sequential case. For later convenience, we show explicitly the $U(1)_Y$ charges for the three components within $(1,{\bf 3},Q_X)$ representation of $\emph{331}$ model, which are given as
\beqa
Q_Y[\Psi_{(1,{\bf 3},Q_X)}]=(\f{1}{6}+Q_X,\f{1}{6}+Q_X,-\f{1}{3}+Q_X)~.
\eeqa

\subsection{The fitting of matter contents into flipped SU(6)}
We propose to partially unify the $\emph{331}$ gauge group into $SU(6)\tm U(1)_K$ gauge group.
The normalized $U(1)_P$ generator within SU(6), which is the (remaining) diagonal generator for $SU(6)$ other than the diagonal ones in $SU(3)_c$ and $SU(3)_L$, can be written as
\beqa
T_P=\f{1}{2\sqrt{3}}(-1,-1,-1,~1,~1,~1)~.
\eeqa
  We can embed the matter contents of $\emph{331}$ model with $\beta=1/\sqrt{3}$ (denoted by their $SU(3)_c\tm SU(3)_L\tm U(1)_P\tm U(1)_K$ quantum numbers) into flipped SU(6) representations for each generation ~\footnote{The case for the partial unification of $\emph{331}$ (with $\beta=\sqrt{3}$) into flipped SU(6) is rather tricky, especially the relevant anomaly cancellation conditions. We will discuss it in our subsequent study. It seems not possible for such a case to unify in ordinary SU(6).}
\beqa
{\bf \bar{6}_{-\f{1}{2}}}&=&U_L^c({\bf \bar{3},1,\f{1}{2\sqrt{3}}})_{\bf -\f{1}{2}}\oplus (L_L, N_L^s) ({\bf 1,\bar{3},-\f{1}{2\sqrt{3}}})_{\bf -\f{1}{2}}~,\nn\\
 {\bf 15}_{0}&=& D_L^c({\bf \bar{3},1,-\f{1}{\sqrt{3}}})_{\bf 0}\oplus (Q_{L},(XD)_L)({\bf 3,3,0})_{\bf 0}
 \oplus ((XL)_L, N_L^c)({\bf 1,\bar{3},\f{1}{\sqrt{3}}})_{\bf 0}~,\nn\\
{\bf \overline{6}}_{\bf \f{1}{2}}
&=&(XD)_L^c({\bf \bar{3},1,\f{1}{2\sqrt{3}}})_{\bf \f{1}{2}}\oplus((XL)_L^c,E_L^c) ({\bf 1,\bar{3},-\f{1}{2\sqrt{3}}})_{\bf \f{1}{2}}~.
\eeqa

The $U(1)_X$ quantum number is related to the corresponding $U(1)_K$ and $U(1)_P$ charges via
\beqa
Q_X=-\f{\sqrt{3}}{3} Q_P+Q_K~,
\eeqa
after the breaking of $SU(6)\tm U(1)_K$ into $SU(3)_c\tm SU(3)_L\tm U(1)_X$.

 It is obvious to see that the  $SU(6)-SU(6)-SU(6)$  anomaly is canceled with two ${\bf \overline{6}}$ representation
and one antisymmetric ${\bf 15}$ representation for each generation, as the anomaly coefficients for SU(6) representations are given by
 \beqa
 {\cal A}({\bf \overline{6}})=-1~,~~~{\cal A}({\bf 15})=2~,
 \eeqa
with $Tr(\{T^a(R),T^b(R)\}T^c(R))=A(R) d^{abc}/2$. The $U(1)_K$ related anomalies are canceled because the $U(1)_K$ quantum numbers for fermions within each generation satisfy
\beqa
6(\f{1}{2}-\f{1}{2})&=&0~,~~~{\rm for}~~~~SU(6)-U(1)_K-U(1)_K~\nn\\
6\[\(\f{1}{2}\)^3-\(\f{1}{2}\)^3\]&=&0~,~~~{\rm for}~~~~~U(1)_K-U(1)_K-U(1)_K
\eeqa
To avoid the gravitational violation of gauge symmetry, the anomaly related to gravity should vanish. It can be seen that the graviton-graviton-$U(1)$ anomaly is canceled in our model, because only the abelian $U(1)_K$ generator is relevant and the $U(1)_K$ charges for the chiral fermions satisfy
\beqa
6(\f{1}{2}-\f{1}{2})&=&0~.
\eeqa
So, we have a possible anomaly free fitting of matter contents in flipped SU(6)
\beqa
{\bf \bar{6}_{-\f{1}{2}}}\supseteq [U_L^c, L_L],~{\bf 15}_{0}\supseteq [Q_L,D_L^c,N_L^c],
~{\bf \overline{6}}_{\f{1}{2}}\supseteq [(XD)_L^c, E_L^c]~,
\eeqa
with $E_L^C\in (1,\bar{3},\f{2}{3})$ of $SU(3)_c\tm SU(3)_L\tm U(1)_X$ and
$N_L^c\in (1,\bar{3},-\f{1}{3})$~\footnote{We note that the fitting of $(XD)_L^c$ and $D_L^C$ can be exchanged (also $(L_L, N_L^s)$ and $((XL)_L, N_L^c)$ ). To ensure the VEV of ${\bf 15}_{0;H}$ is small, such a fitting
\beqa
{\bf \bar{6}_{-\f{1}{2}}}\supseteq [U_L^c, L_L],~{\bf 15}_{0}\supseteq [Q_L,(XD)_L^c,N_L^c],
~{\bf \overline{6}}_{\f{1}{2}}\supseteq [D_L^c, E_L^c]~,
\eeqa
with $E_L^c\in (1,\bar{3},\f{2}{3})$ of $SU(3)_c\tm SU(3)_L\tm U(1)_X$ and
$N_L^c\in (1,\bar{3},-\f{1}{3})$  is not adopted here. We will discuss such alternative choices in our subsequent studies.}.
 We can see that the fillings of ${\bf \overline{6}}_{-\f{1}{2}}$ (containing $U_L^c$ and $L_L$), ${\bf 15}_0$ (containing $Q_L$,$D_L^c$,$N_L^c$) and ${\bf \overline{6}}_{\f{1}{2}}$ (containing $E_L^c$) are similar to that in flipped SU(5).

The fact that the gauge anomaly cancellation of the flipped SU(6) holds for each generation is the reminiscent of the gauge anomaly cancellation conditions of sequential $\emph{331}$ model. Such generation by generation gauge anomaly cancellation conditions will in general not hold for non-sequential $\emph{331}$ model fitting of flipped SU(6) model.

We should briefly comment on the anomaly cancellation conditions in non-flipped versus flipped SU(6) (partial) unification of (non-)sequential $\emph{331}$ models. In the ordinary SU(6) unification of the sequential $\emph{331}$ model, each generation will still be fitted into ${\bf 15\oplus \bar{6}\oplus\bar{6}}$ representations so as that the gauge anomaly cancellation conditions are satisfied for each generation. In the ordinary SU(6) unification of the non-sequential $\emph{331}$ model, the RH-charged leptons $E_{aL}^c\sim {\bf (1, 1, 1)}$ needs to be fitted into ${\bf 20}$ representation of SU(6). The quarks (including the exotic vector-like quarks) and LH leptons can be fitted into ${\bf 15\oplus \bar{6}\oplus\bar{6}}$ for the first two generations, while those of the third generation needs to be fitted into ${\bf 20\oplus 15\oplus15\oplus \bar{6}}$. We can see that the gauge anomaly cancellation conditions for the first two generations in ordinary SU(6) unification of non-sequential $\emph{331}$ model are satisfied for each generation while the third generation needs additional exotic fermions in ${\bf \bar{6}}$ representation to cancel anomaly. Therefore, the anomaly cancellation for non-flipped SU(6) unification of both sequential and non-sequential $\emph{331}$ model always hold generation by generation, even though such conditions for the low energy non-sequential $\emph{331}$ model are satisfied  non-trivially unless contributions from all the three generations are included.

In the flipped SU(6) partial unification of non-sequential $\emph{331}$ model, which will discussed in detail in our future work~\cite{alternative:fsu6}, the RH-charged leptons $E_{aL}^c\sim {\bf (1, 1, 1)}$ can still be kept as SU(6) singlet.  The gauge anomaly for the first two generations, which are given by $15_0\oplus \bar{6}_{1/2}\oplus \bar{6}_{-1/2}\oplus 1_{1}$ will no-longer cancel unless we include the third generation. We need to introduce only one ${\bf 20}$ representation of SU(6) for the third generation, which needs fairly small additional exotic matter contents in contrast to ordinary non-flipped SU(6) unification of non-sequential $\emph{331}$ model.

\subsection{The Higgs sector }
The Higgs fields introduced in our model are responsible for the breaking of $SU(6)\tm U(1)_K$ gauge group, the breaking of the $SU(3)_c\tm SU(3)_L\tm U(1)_X$ gauge group and the mass generations for the matter contents, including the SM matter contents, the exotic vector-like fermions and the sterile neutrinos. The total Higgs sector in our flipped SU(6) model contains the following Higgs fields
\beqa
{\bf 20}_{\f{1}{2};H},~~{\bf \overline{6}}_{\f{1}{2};H},~~{\bf \overline{6}}_{-\f{1}{2};H},~~{\bf 15}_{0;H}, ~~({\bf \overline{105}})^s_{0;H},~~{\bf 21}_{1;H}.
\eeqa
with the gauge symmetry broken by the corresponding VEVs
\beqa
 SU(6)\tm U(1)_K &\xlongrightarrow{~~{\bf 20}_{\f{1}{2};H}~~}{}& SU(3)_c\tm SU(3)_L\tm U(1)_X \nn\\&\xlongrightarrow{{\bf \overline{6}}_{-\f{1}{2};H},({\bf \overline{105}})^s_{0;H},{\bf 21}_{1;H}}{}& SU(3)_c\tm SU(2)_L\tm U(1)_Y\nn\\&\xlongrightarrow{{\bf \overline{6}}_{\f{1}{2};H},{\bf 15}_{0;H},{\bf 21}_{1;H}}& SU(3)_c\tm U(1)_Y,
\eeqa
To break the flipped SU(6) into $SU(3)_c\tm SU(3)_L\tm U(1)_X$, we introduce a ${\bf 20}_{\f{1}{2}}$ representation Higgs with its decomposition in terms of $SU(3)_c\tm SU(3)_L\tm U(1)_P\tm U(1)_K$ quantum numbers
\beqa
{\bf {20}}_{\f{1}{2};H}=(1,1,-\f{3}{2\sqrt{3}})_{\f{1}{2};H}\oplus(1,1,\f{3}{2\sqrt{3}})_{\f{1}{2};H}\oplus
(3,\bar{3},-\f{1}{2\sqrt{3}})_{\f{1}{2};H}  \oplus(\bar{3},3,\f{1}{2\sqrt{3}} )_{\f{1}{2};H}~,\nn\\
\eeqa
with
\beqa
Q_X: (1,1,\f{3}{2\sqrt{3}})_{\f{1}{2};H}=0~,~~Q_X:(1,1,-\f{3}{2\sqrt{3}})_{\f{1}{2};H}=-1~.
\eeqa
The $(1,1,\f{3}{2\sqrt{3}})_{\f{1}{2};H}$ component of ${\bf 20}_{\f{1}{2};H}$ can acquire a vacuum expectation value (VEV)
$\langle {\bf {20}_{\f{1}{2};H}}\rangle= M_X$ to break the flipped SU(6) into $SU(3)_c\tm SU(3)_L\tm U(1)_X$.

To break the residue $SU(3)_c\tm SU(3)_L\tm U(1)_X$ gauge symmetry into SM and generate masses for the SM matter contents, we need to introduce additional Higgs fields ${\bf \overline{6}}_{\f{1}{2};H}$, ${\bf 15}_{0,H}$ and ${\bf \overline{6}}_{-\f{1}{2};H}$ with their decompositions in terms of $SU(3)_c\tm SU(3)_L\tm U(1)_P\tm U(1)_K$ quantum number
\beqa
{\bf \overline{6}}_{\f{1}{2};H}&=&
(\bar{3},1,\f{1}{2\sqrt{3}})_{\f{1}{2};H}\oplus (1,\bar{3},-\f{1}{2\sqrt{3}})_{\f{1}{2};H},\nn\\
{\bf \overline{6}}_{-\f{1}{2};H}&=&
(\bar{3},1,\f{1}{2\sqrt{3}})_{-\f{1}{2};H}\oplus (1,\bar{3},-\f{1}{2\sqrt{3}})_{-\f{1}{2};H},\nn\\
{\bf 15}_{0;H}&=&(\bar{3},1,-\f{1}{\sqrt{3}})_{0;H}\oplus (3,3,0)_{0;H}
 \oplus (1,\bar{3},\f{1}{\sqrt{3}})_{0;H}~.
\eeqa
The Yukawa couplings for the matter contents can be written as
\beqa
{\cal L}&\supseteq& -\sum\limits_{a,b=1}^3Y_{U;ab}{\bf \overline{6}}^a_{-\f{1}{2}}{\bf 15}^b_0{\bf \overline{6}}_{\f{1}{2};H}
-\sum\limits_{a,b=1}^3Y_{E;ab} {\bf \overline{6}}^a_{-\f{1}{2}}{\bf \overline{6}}^b_{\f{1}{2}} {\bf 15}_{0,H}
-\sum\limits_{a,b=1}^3Y_{D,N;ab} {\bf 15}^a_0{\bf 15}^b_0{\bf {15}}_{0;H}\nn\\
&-&\sum\limits_{a,b=1}^3Y_{XD;ab}{\bf \overline{6}}^a_{\f{1}{2}}{\bf 15}^b_0{\bf \overline{6}}_{-\f{1}{2};H},
\label{Yukawa:fsu6}
\eeqa
with $'a,b'$ the family indices.
The low energy Higgs fields in $SU(3)_c\tm SU(3)_L\tm U(1)_X$ models contain
the $(1,\bar{3},-\f{1}{2\sqrt{3}})_{\f{1}{2};H}\in {\bf \overline{6}}_{\f{1}{2};H}$ Higgs field
(corresponds to $H_1(1,\bar{3},\f{2}{3})$ that contains doublet $H_u$ Higgs field in 2HDM),
the $(1,\bar{3},-\f{1}{2\sqrt{3}})_{-\f{1}{2};H}\in {\bf \overline{6}}_{-\f{1}{2};H}$ Higgs field
(corresponds to $H_2(1,\bar{3},-\f{1}{3})$)
and the $(1,\bar{3},\f{1}{\sqrt{3}})_{0;H}\in {\bf 15}_{0;H}$ Higgs field (corresponds to $H_3(1,\bar{3},-\f{1}{3})$ that contains doublet $H_d$ Higgs field in 2HDM), which are just needed to
generate properly the masses for matter contents. As the ${\overline{6}}_{-\f{1}{2};H}$ Higgs is not responsible for the masses generations of SM matter contents, the simplest choice to break $SU(3)_c\tm SU(3)_L\tm U(1)_X$ to SM is to adopt the VEV for its triplet component $H_2(1,\bar{3},-\f{1}{3})$, which can be decomposed with the corresponding SM gauge quantum number
\beqas
H_2(1,\bar{3},-\f{1}{3})=H^\pr(1,\bar{2},\f{2}{3})\oplus N_H^\pr(1,1,0)~,
\eeqas
along the $(1,1,0)$ direction (in terms of SM gauge quantum number) as $\langle H_2\rangle=M_{331}$. The VEVs of the relevant Higgs fields can be written as
\beqa
\langle H_1\rangle=\sqrt{2}\(\bea{c}v_u\\0\\0\eea\),~~\langle H_2\rangle=\sqrt{2}\(\bea{c}0\\0\\M_{331}\eea\),~~\langle H_3\rangle=\sqrt{2}\(\bea{c}v_d\\0\\0\eea\),
\eeqa
with the VEVs of $H_1$ and $H_3$ trigger the breaking of the SM electrweak symmetry into $U(1)_Q$.

It can be seen from the Yukawa couplings in eq.(\ref{Yukawa:fsu6}) that the ${\bf \overline{6}}^a_{\f{1}{2}}{\bf 15}^b_0{\bf \overline{6}}_{-\f{1}{2};H}$ term will leads to
\beqa
\[(1,\bar{3},-\f{1}{2\sqrt{3}})_{\f{1}{2}}\otimes(1,\bar{3},\f{1}{\sqrt{3}})_{0}
\otimes (1,\bar{3},-\f{1}{2\sqrt{3}})_{-\f{1}{2};H}\]&\supseteq& \[(XL)_L\otimes (XL)_L^{c}\otimes N_H^\pr(1,1,0)\]~,\nn\\
\[(3,3,0)_{0}\otimes (\bar{3},1,\f{1}{2\sqrt{3}})_{\f{1}{2}}\otimes (1,\bar{3},-\f{1}{2\sqrt{3}})_{-\f{1}{2};H}\]&\supseteq& \[(XD)_L \otimes (XD)_L^c \otimes N_H^\pr(1,1,0)\]~,\nn
\eeqa
which will generate Dirac mass $Y_{XD} M_{331}$ for vector-like heavy extra leptons
$(XL)_L,(XL)^c_L$ and vector-like heavy quarks $(XD)_L,(XD)_L^c$.

 Experimental measurements for the square of the mass differences~\cite{neutrino:report} for neutrinos indicate that the heaviest neutrino mass should be of order $10^{-2} {\rm eV}$. Such tiny neutrino masses can either be Dirac type or be Majorana type from dim-5 Weinberg operator. We should note that it is not possible to adopt only the Dirac type masses for neutrinos because the Yukawa terms involving $Y_{U;ab}$ generate identical masses for both the up-type quark masses and Dirac-type neutrino masses at the flipped SU(6) breaking scale $M_X$. Large hierarchy between the up-type quark masses and tiny Dirac type neutrino masses can not be generated by pure renormalization group equation (RGE) effects, that is, by RGE evolution from $M_{X}$ to $M_Z$. Tiny Majorana neutrino masses from dim-5 Weinberg operator can be UV completed to various mechanisms, for example, the Type-I seesaw mechanism, which can be used to generate tiny neutrino masses after introducing additional Majorana mass terms for RH-neutrinos $N_L^c$. Bare Majorana mass terms are not allowed because the RH-neutrinos are fitted into non-singlet ${\bf 15}_0$ representations of $SU(6)\tm U(1)_K$.
So, such Majorana mass terms for RH neutrinos can be only generated by a new term involving certain new Higgs field that couples to RH neutrinos. From the production of ${\bf 15}_0$ representation
\beqa
{\bf 15}\otimes {\bf 15}={\bf \overline{15}}\oplus {\bf 105}^s\oplus{\bf 105}^a~,
\eeqa
we can see that the proper choice is ${\bf \overline{105}_0^s}$, which is decomposed as
\small
\beqa
({\bf \overline{105}})_{0;H}^s={\bf (1,{6},-\f{2}{\sqrt{3}})_0\oplus({6},1,\f{2}{\sqrt{3}})_0\oplus(8,\bar{3},\f{1}{\sqrt{3}})_0
\oplus(\bar{3},{8},-\f{1}{\sqrt{3}})_0\oplus(6,6,0)_0\oplus({3},{3},0)_0}~,\nn
\eeqa
\normalsize
in terms of $SU(3)_c\tm SU(3)_L\tm U(1)_P\tm U(1)_K$ quantum numbers. The term responsible for the generation of Majorana neutrino masses can be written as
\beqa
{\cal L}\supseteq Y^m_{ab}{\bf 15}_{0}^a{\bf 15}_0^b ({\bf \overline{105}})^s_{0;H}\supseteq (1,\bar{3},\f{1}{\sqrt{3}})^a_{0}\otimes(1,\bar{3},\f{1}{\sqrt{3}})^b_{0}\otimes(1,{6},-\f{2}{\sqrt{3}})_{0;H}~,
\eeqa
which will generate Majorana masses for RH neutrinos after the $(1,{6},-\f{2}{\sqrt{3}})_{0;H}$ component of ${\bf \overline{105}}_s$ develops a VEV along the $(1,1,0)$ direction (in terms of SM quantum number).
 As $\langle {\bf \overline{105}}_s \rangle=M_S$ will also break the $SU(3)_c\tm SU(3)_L\tm U(1)_X$ gauge symmetry, we require that $M_S\lesssim M_{331}$. The neutrino masses can be given by
 \beqa
 {\cal M}_\nu=\(\bea{cc}0& {\cal M}_{\nu;D}^{T}\\{\cal M}_{\nu;D} & Y^m M_S\eea\)~.
\label{mass:neutrino} \eeqa
The natural up-type quark masses, which are also the typical Dirac-type neutrino masses, are given by $m_U\simeq Y^U v_u\simeq {\cal M}_{\nu;D}\sim {\cal O}(10^2)$ GeV. To obtain tiny neutrino masses of order $10^{-2}$ eV with the seesaw mechanism
\beqa
M_\nu\simeq \f{M_{\nu;D} M_{\nu;D}^T}{Y^m M_S}\sim 5\tm 10^{-2} {\rm eV}~,
\eeqa
the $331$ breaking scale $M_{331}$ is constrained to lie naturally at about $10^{14}$ {\rm GeV} for $Y^m\sim {\cal O}(1)$, otherwise the generated neutrino masses should be much larger than $10^{-2}$ {\rm eV}. The bounds on $M_{331}\sim 10^{14} {\rm GeV}$ from neutrino masses can be relaxed to $M_{331}\gtrsim 10^{14} {\rm GeV}$ if the coupling $Y^m$ can be much smaller than identity. We should note that constraints on the scale of $M_{331}$ from neutrino masses can be relaxed if a mixed type I+ II seesaw mechanism  is used for neutrino masses generations, within which a small VEV for an additional ${\bf 21}_{1;H}$ representation Higgs field along the $SU(2)_L$ triplet direction is needed. We will discuss such a possibility shortly after. On the other hand, it will be clear soon that successful gauge couplings unification for $g_{3c}$ and $g_{3L}$ requires the $M_{331}$ scale to be higher than $10^{16}$ GeV. Such a bound can be relaxed unless certain additional colored Higgs field lies of order $M_{331}$ scale.
 Given the neutrino masses, the Yukawa coupling involved can be defined in terms
of the physical neutrino parameters, up to an orthogonal complex matrix $R$~\cite{Yukawa:neutrino},
\beqa
Y^U\approx {\sqrt{2}}\f{i}{v_u}\sqrt{\hat{M}_R} R\sqrt{\hat{m}_\nu} V_{PMNS}^\da~,
\eeqa
where $\hat{m}_\nu$, $\hat{M}_R$ being the diagonal matrices for the light and heavy neutrino masses, and $V_{PMNS}$ being the PMNS lepton mixing matrix. In our case, the neutrino hierarchical spectrum can either be normally ordered (NO) or inversely ordered (IO), depending on the Yukawa parameters introduced in the theory.

 If we adopt non-renormalizable Weinberg operator
 \beqa
 \Delta {\cal L}_{\rm Weinberg}= \f{y_{ab}^\nu}{M} \( \bf{\bar{6}_{-1/2;a}\bar{6}_{-1/2;b}}\) \(\bf{\bar{6}_{1/2,H}\bar{6}_{1/2,H}}\)~,
 \eeqa
to generate tiny neutrino mass for flipped SU(6) without specifying its concrete UV completion model, the previous lower bound on $M_{331}$ from neutrino mass generation can be relaxed. Here $M$ denotes the scale of the heavy modes, which are integrated out and responsible for the generation of Weinberg operator. As the non-renormalizability of the Weinberg operator requires $M$ to be larger than the flipped SU(6) breaking scale $M_X$, we thus obtain an upper bound for $M_X$ with $M_X < 10^{14}$ GeV in this case. To be consistent, we need to ensure that such a constraint is satisfied for the choice of $M_{331}$ scale and the Higgs contents of the low energy $\emph{331}$ model. We leave the numerical discussions of this possibility in our future work.

The new sterile neutrino component $N^s_L$ within ${\bf \bar{6}_{-\f{1}{2}}}$ can also obtain masses after EWSB, which couples to the $(XL)_L$ and $(XL)_L^c$ components via the Yukawa coupling terms involving $Y_U$ and $Y_E$
\beqa
\[(1,\bar{3},-\f{1}{2\sqrt{3}})_{-\f{1}{2}}\otimes(1,\bar{3},\f{1}{\sqrt{3}})_{0}
\otimes (1,\bar{3},-\f{1}{2\sqrt{3}})_{\f{1}{2};H}\]&\supseteq& \[N^S_L\otimes (XL)_L\otimes H_u \]~,\nn\\
\[(1,\bar{3},-\f{1}{2\sqrt{3}})_{-\f{1}{2}}\otimes (1,\bar{3},-\f{1}{2\sqrt{3}})_{\f{1}{2}}
\otimes (1,\bar{3},\f{1}{\sqrt{3}})_{0;H}\]&\supseteq& \[N^S_L\otimes (XL)_L^{c}\otimes H_d \]~.
\eeqa
So the mass matrix for the new sterile neutrinos can be given by
\beqa
M_{S}^\pr\equiv \(\bea{ccc}0&{\cal M}_U^T &{\cal M}_E^T \\{\cal M}_U &0&Y_{XD} M_{331}\\ {\cal M}_E &Y_{XD} M_{331} &0 \eea\)~,
\label{SNeutrino}
\eeqa
in the basis of $N_L^s, N_{(XL)}, N_{(XL)^c} $. Here $N_{(XL)}$, $N_{(XL)^c}$ denote the neutral components within $(XL)_L$ and $(XL)^c_L$, respectively. The $M_U$, $M_E$ scales lie typically at the  up-type quark mass scales ${\cal O}(10^2)$~{\rm  GeV} and charged lepton mass scales ${\cal O}(1)$~{\rm  GeV}, respectively. After diagonalizing the mass matrix, we can obtain that the mass scale for the lightest new sterile neutrino is
\beqa
m_S\sim \f{{\cal M}_U {\cal M}_E }{Y_{XD} M_{331}}\sim 10^{-3}~{\rm eV}~,
\label{sterile:mass}
\eeqa
for $Y_{XD}\sim {\cal O}(1)$, which can contribute to additional light effective degrees of freedom $\Delta g_*$ at the BBN era and cause cosmological difficulties. Therefore, we should try to push heavy such new sterile neutrinos, for example, by choosing unnaturally small $Y_{XD}$.
 An interesting solution to such a problem without unnatural parameters is to introduce additional Majorana type masses for $N_L^s$. We can introduce new ${\bf 21}_{1;H}$ representation Higgs field, which has the following decomposition
\beqa
{\bf 21}_{1;H}=({6},1,-\f{1}{\sqrt{3}})_{1;H}\oplus (3,3,0)_{1;H}
 \oplus (1,{6},\f{1}{\sqrt{3}})_{1;H}~,
\eeqa
in terms of $SU(3)_c\tm SU(3)_L\tm U(1)_P\tm U(1)_K$ quantum numbers and the relevant Yukawa coupling is
\beqa
{\cal L}\supseteq -y_{S;ab}{\bf \overline{6}}^a_{-\f{1}{2}}{\bf \overline{6}}^b_{-\f{1}{2}} {\bf {21}}_{1;H}~.
\eeqa
When the $(1,{6},\f{1}{\sqrt{3}})_{1;H}$ component of ${\bf 21}_{1;H}$ develops a VEV with $\langle{\bf 21}_{1;H}\rangle=M_{S^\pr}\sim M_{331}$ along the $(1,1,0)$ direction (in terms of SM quantum number), Majorana mass term can be generated for $N_L^S$. With new contribution $(M_{S}^\pr)_{11}\sim M_{331}$ in eq.(\ref{SNeutrino}), the eigenvalues of $M_{S}^\pr$ all lie at the $M_{331}$ scale and will not cause cosmological difficulties. On the other hand, if $M_{331}\gg\langle{\bf 21}_{1;H}\rangle\gtrsim 0.1~{\rm }keV$ (or choosing $Y_{XD}\lesssim 10^{-6}$ for the first solution), the lightest sterile neutrinos with masses of order $\langle{\bf 21}_{1;H}\rangle$ can act as a fermionic dark matter candidate, which also satisfy the Tremaine-Gunn (TG) bound~\cite{TG:fermionicDM}.

Besides, if the $(1, {3},1)$ direction (in terms of SM quantum number) of $(1,{6},\f{1}{\sqrt{3}})_{1;H}$ component within ${\bf 21}_{1;H}$ Higgs field also develops a small triplet VEV (which also breaks the SM electroweak gauge symmetry), ordinary LH neutrinos of SM can also acquire Majorana masses so as that a mixed type I+II seesaw mechanism can be applied to the non-sterile neutrino sector. With a small non-vanishing $({\cal M}_\nu)_{11}$ component for the the mass matrix (\ref{mass:neutrino}), the $331$ breaking scale $M_{331}$ can be much lower than $10^{14}$ GeV with large fine-tuning among the type-I and type-II seesaw contributions for the neutrino masses. For example, the choice of $M_{331}\sim 10^3~{\rm GeV}$ requires ${\cal O}(10^{-11})$ fine tuning for both contributions to get tiny neutrino masses of order $10^{-2}{\rm eV}$. However, our numerical results indicate that successful gauge couplings unification for $g_{3c}$ and $g_{3L}$ still requires large $M_{331}$ scale (larger than $10^{12}$ GeV) in the case with small $SU(2)_L$ triplet VEV, unless we keep some additional Higgs field as light as $M_{331}$ scale, for example, keeping the ${\bf (\bar{3},8,-\f{1}{\sqrt{3}})_0}$ Higgs field within $\overline{105}^s$ to lie at $M_{331}$ scale.

It is worth to note that the Higgs field in ${\bf 6, 15,20}$ representation of SU(6) can be generated at the Kac-Moody level one~\cite{string:GUT}. The large representation ${\bf \overline{105}^s}$ and ${\bf 21}$ Higgs fields from higher Kac-Moody level can in fact be replaced by double ${\bf 6}$ or $\bar{15}$ Higgs fields, similar to that appeared in the non-renormalizable Weinberg operators.

\section{\label{sec-3} Gauge Coupling Unification}
The GUT symmetry breaking chain is given by
\beqa
SO(12)/E_6&\stackrel{M_{G}}{\longrightarrow}& SU(6)\tm U(1)_K \stackrel{M_{X}}{\longrightarrow} SU(3)_c\tm SU(3)_L\tm U(1)_X \nn\\&\stackrel{M_{331}}{\longrightarrow}& SU(3)_c\tm SU(2)_L\tm U(1)_Y\stackrel{M_{Z}}{\longrightarrow} SU(3)_c\tm U(1)_Y,
\eeqa
with the partial unification $SU(6)\tm U(1)_K$ gauge group can be further unified into a simple $SO(12)$ group ( or $E_6$ group via intermediate $SU(2)\tm SU(6)$ step).
The relations among the $U(1)$ generators
\beqa
Q_X=-\f{\sqrt{3}}{3} Q_P+Q_K,~~~~~Q_Y=\f{1}{\sqrt{3}}T_8 + Q_X~,
\eeqa
leads to the relations for the relevant gauge couplings
\beqa
\f{1}{g_X^2}=\f{1}{3}\f{1}{g_P^2}+\f{1}{g_K^2}~,~~ ~\f{1}{g_Y^2}=\f{1}{3}\f{1}{g_{3L}^2}+\f{1}{g_X^2}~,
\eeqa
holding at the $SU(6)\tm U(1)_K$ breaking scale $M_X$ (for $1/g_X^2$) and the $M_{331}$ scale (for $1/g_Y^2$), respectively.  If we fit the flipped $SU(6)$ gauge couplings within $SO(12)$, the coupling $g_K$ should be normalized into canonical $g_{\tl{K}}$ with $g^2_K=g_{\tl{K}}^2/3$. It should be noted that the charge quantization can only be explained in the framework of $SO(12)$ or $E_6$ GUT instead of our partial unification scheme. The $U(1)_K$ charge assignments in our intermediate partial unification $SU(6)\tm U(1)_K$ model are still not quantized, which are constrained only by gauge anomaly cancellation conditions.

The one-loop beta function for the couplings are given by
\beqa
\f{d}{dt}g_i=\f{1}{16\pi^2} b^0_i g_i^3~,
\eeqa
with
\beqa
b_0^i=-\f{11}{3}C(G_i)+\f{2}{3}\sum\limits_{fermion} T(r_f^i)+\f{1}{3}\sum\limits_{scalar} T(r_s^i)~,
\eeqa
for Weyl fermions in ${\bf r}_f^i$ representation and complex scalars in ${\bf r}_s^i$ representations.

In SUSY SU(5) GUT model, the doublet Higgs field that responsible for electroweak symmetry breaking should be much lighter than the colored triplet Higgs field  so as that the dim-5 operator induced proton decay mode suppressed by the triplet Higgs mass can still be consistent with current proton decay bounds. There are many proposals to deal with such doublet-triplet (D-T) splitting problem, such as the missing partner mechanism~\cite{missing-partner}, complicated version of sliding singlet mechanism in SU(6) extension~\cite{sliding-singlet}, missing VEV in SO(10)~\cite{missing-VEV}, pseudo Nambu-Goldstone bosons~\cite{PNG} etc. In missing partner mechanism, the color-triplet Higgs fields can couple with other colored fields to acquire large masses, whereas the doublet Higgs fields lack such partners so as that they can still be light. Missing partner mechanism can be elegantly realized in flipped SU(5) GUT model, which does not require adjoint or larger Higgs representations and can be seen as a virtue of flipped SU(5). We should note that the missing partner mechanism can also be realized in our flipped SU(6) model (see the discussions in the appendix~\ref{appendix-1}).

In our following discussions, we assume that the splitting among the colored/uncolored Higgs fields can be successfully realized and we do not specify the origins of such splitting, for example, by missing partner mechanism or by orbifold projection (see appendix~\ref{appendix-2}).  Therefore, rendering the colored Higgs fields to lie near the SU(6) breaking scale, the Higgs sector for the low energy $\emph{331}$ model contains
\beqa
H_1(1,\bar{3},\f{2}{3})\ni H_u,~~H_2(1,\bar{3},-\f{1}{3})\ni H_d,~~ H_3(1,\bar{3},-\f{1}{3}),~~H_S^i(1,6,\f{2}{3})~(i=1,2).
\eeqa
We have the following coefficients for $\emph{331}$ model
\beqa
(b_3^c,b_3^L,b_1^X)=(-5,-\f{17}{6},\f{94}{9}).
\eeqa
Upon the $M_X$ scale, where $\al_{3c}(M_X)=\al_{3L}(M_X)=\al_6(M_X)\equiv\al_X$, the flipped $SU(6)$ fields include the matter contents for three generations ${\bf \bar{6}}_{-\f{1}{2};i}$,${\bf \bar{6}}_{\f{1}{2};i}$,${\bf 15}_{0;i}$ and the Higgs fields ${\bf 20}_{-\f{1}{2};H}$,${\bf 21}_{1;H}$,${\bf \overline{105}}_{0;H}$,${\bf \overline{6}}_{\f{1}{2};H}$, ${\bf 15}_{0,H}$ and ${\bf \overline{6}}_{-\f{1}{2};H}$. The beta functions for $SU(6)$ and $U(1)_K$ are given by
\beqa
(b_6,b_K)=(-\f{38}{3},\f{47}{3})~,
\eeqa
which, after normalization into $SO(12)$, gives
\beqa
(b_6,b_{K^\pr})=(-\f{38}{3},\f{47}{9})~.
\eeqa
After the breaking of $\emph{331}$ gauge group into $SU(3)_c\tm SU(2)_L\tm U(1)_Y$ at about the $M_{331}$ scale, the theory  reduce to two Higgs doublet model (requiring $M_{331}\gtrsim 10^{14}$ {\rm GeV}) or two Higgs doublet Model plus an $SU(2)_L$ triplet (with relaxed $M_{331}$ scale, for example, at TeV scale, although large fine tuning is needed).

The relevant beta functions are given by
\bit
\item Case I: two Higgs doublet model~(2HDM)
\beqa
(b_3,b_2,b_Y)=(-7,-3,7)~.
\eeqa
\item Case II: 2HDM plus an $SU(2)_L$ triplet
\beqa
(b_3,b_2,b_Y)=(-7,-\f{7}{3},8)~.
\eeqa
\eit
Note that we do not have the $\f{3}{5}$ factor for $b_Y$ because we do not normalize the $g_Y$ couplings within $SU(5)$. We adopt the following inputs at $M_Z$ scale~\cite{PDG2022}
\beqa
\al_{em}^{-1}=127.951\pm 0.009~,~~\sin^2\theta_W=0.23129 \pm 0.00033~,~\al_s(M_Z)= 0.1185\pm 0.0016~,\nn
\eeqa
to obtain the central values~\cite{Degrassi:2012ry}
\beqa
\al_2^{-1}(M_Z)=29.594~,~~\al_Y^{-1}(M_Z)=98.357~, ~~g_3(m_t) = 1.3075~.
\eeqa

 We can calculate the relevant $SU(6)\tm U(1)_K$ breaking scale $M_X$ for various low energy cases, after specifying the matter and Higgs contents of the low energy $\emph{331}$ models and their corresponding low energy  models at the electroweak scale (in our case, the 2HDM or 2HDM plus an $SU(2)_L$ triplet). Given an $M_{331}$ scale, the $SU(6)\tm U(1)_K$ breaking scale and the corresponding $SU(5)$ gauge coupling can be obtained numerically, using the corresponding beta functions for the gauge couplings given in previous discussions. In our partial unification model, the $M_X$ scale is defined as the intersection scale of the RGE evolution trajectories for $SU(3)_c$ and $SU(3)_L$ gauge couplings. The $U(1)_K$ coupling strength at $M_X$ can be obtained by the combinations of $U(1)_X$ coupling and the coupling of gauge field corresponding to the diagonal $U(1)_P$ generator within SU(6), which is just the SU(6) gauge coupling strength at $M_X$. The RGE evolution trajectory of $U(1)_K$ gauge coupling upon $M_X$ will eventually intersect/unify with that of $SU(6)$ gauge coupling at the $SO(12)/E_6$ unification scale (or at the string scale $M_{str}$ with gravity).

\begin{figure}[!ht]
\begin{center}
\includegraphics[width=2.8in, height=2.8in]{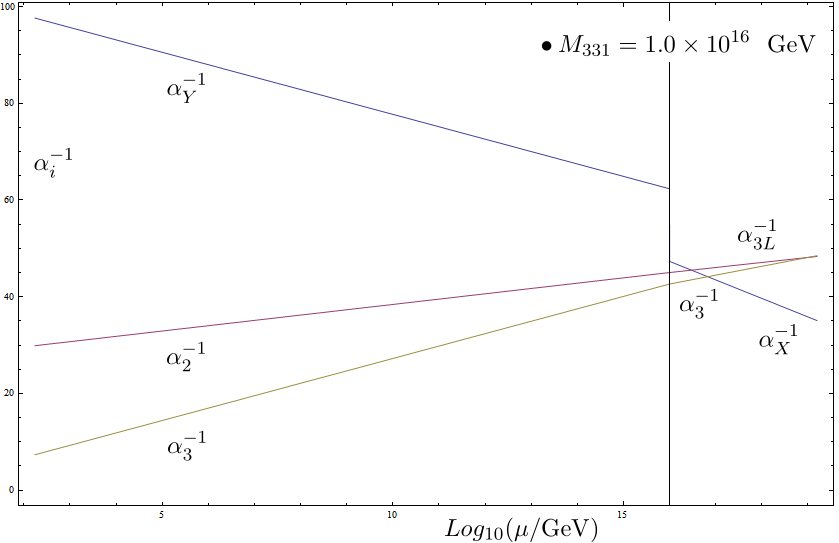}
\includegraphics[width=2.8in, height=2.8in]{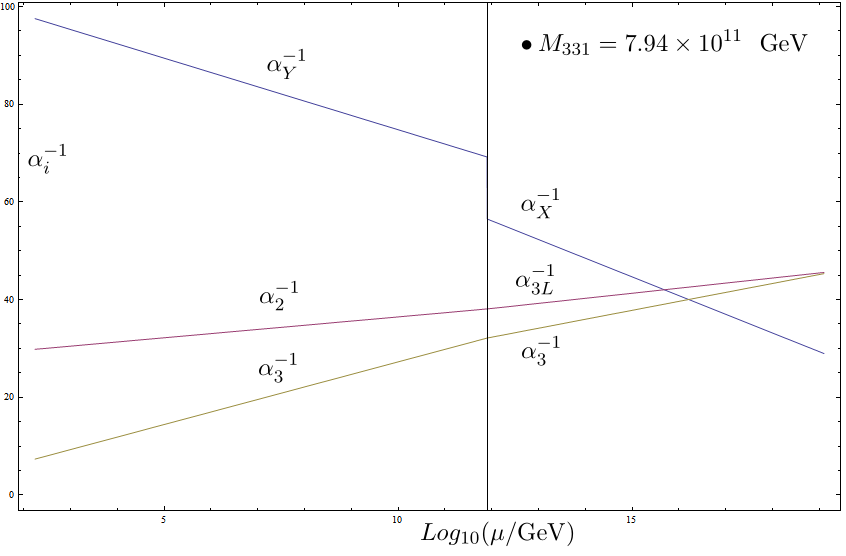}\\
\includegraphics[width=2.8in, height=2.8in]{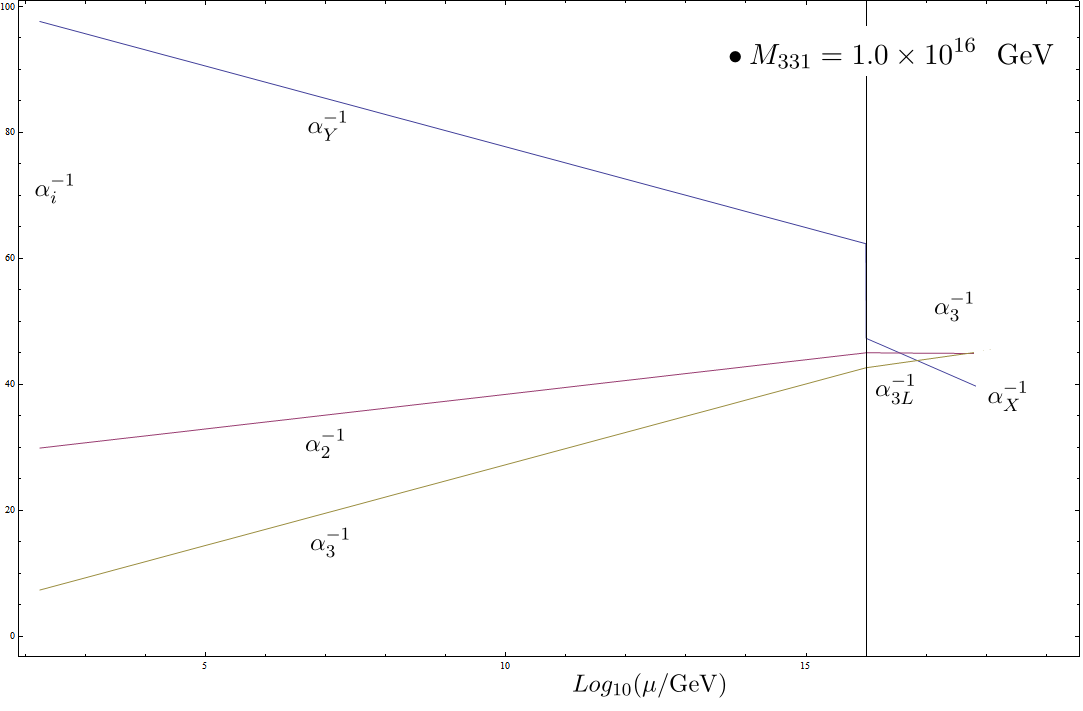}
\includegraphics[width=2.8in, height=2.8in]{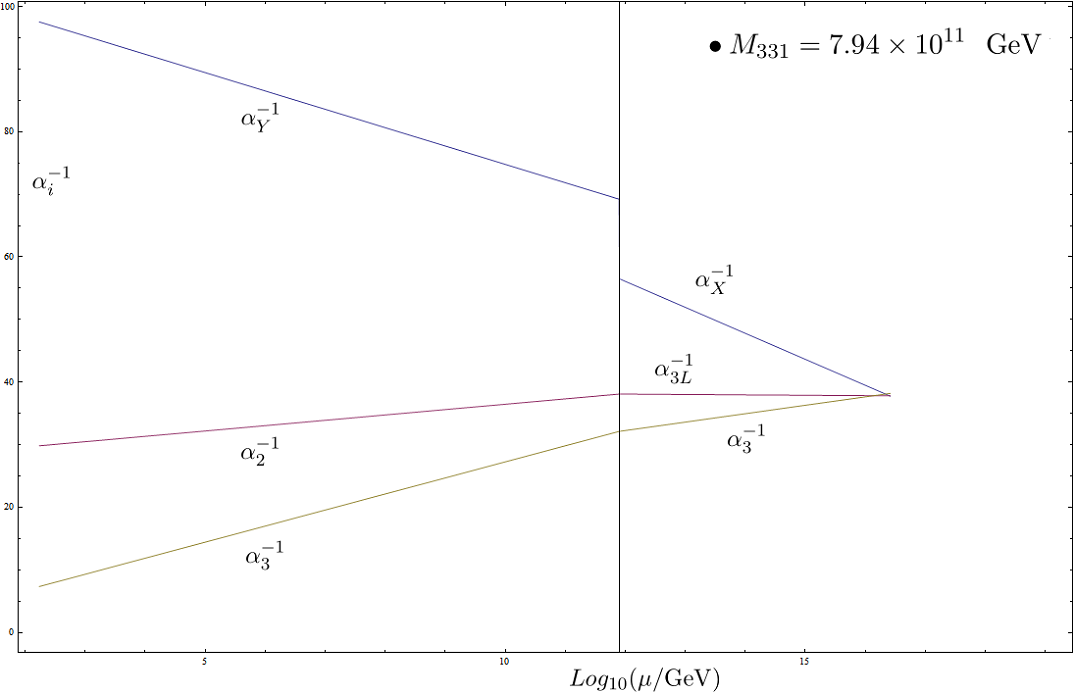}\\
\end{center}
\vspace{-.5cm}
\caption{The RGE evolutions of the gauge couplings for the $\emph{331}$ models are shown for scenarios with 2HDM below $M_{331}$ (case I, left panels) and 2HDM plus $SU(2)_L$ triplet Higgs below $M_{331}$ (case II, right panels) without light ${\bf \tl{H}_{3,8}}$ Higgs, respectively.
With the $\emph{331}$ symmetry breaking scale $M_{331}=10^{16}$ GeV (left panels) and $M_{331}=7.94\tm 10^{11}$ GeV (right panels), the $SU(3)_L$ and $SU(3)_c$ gauge couplings can be unified into the flipped $SU(6)$ GUT model at the scale $M_X=10^{19.03}$  GeV (left panel) and  $M_X=10^{19.13}$  GeV (right panel), respectively. The upper (and lower) panels correspond to the cases without (and with) the surviving $M_{331}$ scale ${\bf \tl{H}_{3,8}}$ Higgs field, respectively.
}
\label{fig1}
\end{figure}

 We randomly scan the values of $M_{331}$ within the ranges that are compatible with the lower bound from  neutrino masses generation and the upper Planck scale bound. Our numerical results indicate that the flipped $SU(6)$ unification of $\emph{331}$ model can indeed be possible. In fig.\ref{fig1}, we show the RGE evolutions of the gauge couplings for scenarios with the low energy theory as 2HDM below $M_{331}$ scale (case I, left panel) and 2HDM plus $SU(2)_L$ triplet Higgs below $M_{331}$ scale (case II, right panel), respectively. The corresponding $\emph{331}$ symmetry breaking scale for the benchmark scenarios are $M_{331}=10^{16}$ GeV (left panel)  and $M_{331}=7.94\tm 10^{11}$ GeV (right panel), respectively. The corresponding flipped $SU(6)$ unification scales are $M_X=10^{19.03}$ GeV with $\al_6^{-1}(M_X)\approx 48.05$ (left panel) and $M_X=10^{19.13}$ GeV with $\al^{-1}_6(M_X)\approx 45.12$ (right panel). On the other hand, requiring the unification scales to lie below the Planck scale constrains $M_{331}\gtrsim 10^{15.9}$ GeV for case I and $M_{331}\gtrsim 10^{11.8}$ GeV for case II. Larger $M_{331}$ scale will lead to smaller flipped $SU(6)$ GUT scale $M_X$. Besides, the requirement that the $SU(2)_L$ coupling (within $SU(3)_L$) and $SU(3)_c$ coupling should not intersect below $M_{331}$ scale, that is, $M_X\gtrsim M_{331}$, set an upper bound for $M_{331}$ scale, which requires it to lie below  $10^{17.6}$ GeV for case I and below $10^{15.3}$ GeV for case II.

%%%%%%%%%%%%%%%%%%%%%%%%%%Fig1%%%%%%%%%%%%%%%%%%
%%%%%%%%%%%%%%%%%%%%%%%%%%%%%%%%%%%%%%%%%%%%%%%%%%%%%
We should note that it is possible to push down the $M_{331}$ scale without spoiling successful flipped $SU(6)$ GUT by keeping the ${\bf (\bar{3},8,-\f{1}{\sqrt{3}})_0}$ Higgs field ${\bf \tl{H}_{3,8}}$ (corresponding to the $\emph{331}$ quantum number ${\bf (\bar{3},8,\f{1}{{3}})}$) as light as the $M_{331}$ scale, for example, by choosing proper boundary condition in the orbifolding projection. In this case, the gauge beta functions change into
 \beqa (b_3^c,b_3^L,b_1^X)=(-\f{11}{3},\f{1}{6},\f{34}{3}),\eeqa upon the $M_{331}$ scale with the relative running $b_{32}\equiv b_{3c}-b_{3L}$ increasing to $\f{23}{6}$.
  The evolutions of the gauge couplings are also shown in the lower panels in fig.\ref{fig1}.
  We show in Tab.\ref{bm1} the flipped SU(6) GUT scales and corresponding $\al_6^{-1}(M_X)$ values for some benchmark points in case I/case II with and without the surviving light ($M_{331}$ scale) ${\bf \tl{H}_{3,8}}$ Higgs, respectively. The upper bounds $10^{17.6}$ GeV for case I (and  $10^{15.3}$ GeV for case II) on $M_{331}$ scale also apply here to guarantee $M_X\gtrsim M_{331}$. On the other hand,
 the requirement $M_{Pl}\gtrsim M_{X}$, that is, the unification scale should lie below the Planck scale, will not give interesting constrains on $M_{331}$ for both case I and case II with light ($M_{331}$ scale) ${\bf \tl{H}_{3,8}}$ Higgs.

%%%%%%%%%%%%%%%%%%%%%%%%%%%%%%%%%%%%%
\begin{table}[htbp]
 \caption{We list the flipped SU(6) GUT scales $\log_{10}\(\f{M_X}{\rm GeV}\)$ and corresponding $\al_6^{-1}(M_X)$ values for some benchmark points for various $M_{331}$ values in case I and case II  with/without light ${\bf \tl{H}_{3,8}}$ Higgs, respectively. The $'\setminus'$ symbols, followed by reasons that denoted by $R_i$, indicate that either the unification scale $M_X$ lies upon the Planck scale (denoted by $R_1$), or $g_{3L}$ and $g_{3c}$ cannot unify (denoted by $R_2$), or the bound on $M_{331}$ from neutrino mass generations is not satisfied for case I (denoted by $R_3$).
  }
 \begin{tabular}{||c|c|c|c|c||}
 \hline
 \hline
\multirow{2}{*}{($\log_{10}\(\f{M_X}{\rm GeV}\),\al_6^{-1}(M_X)$)} &\multicolumn{4}{|c|}{$M_{331}/{\rm GeV}$}\\
 \cline{2-5}
  &$3.16\tm 10^{3}$ &$1.0\tm 10^{11}$ &$5.0\tm 10^{14}$&$3.16\tm 10^{16}$  \\
 \hline\hline
 Case I without ${\bf \tl{H}_{3,8}}$&$\setminus; R_3$&$\setminus;R_3$&$\setminus;R_1$&$(18.61,47.77)$\\
 \hline
 Case II without ${\bf \tl{H}_{3,8}}$&$\setminus; R_1$&$\setminus;R_1$&$(16.18,42.02)$&$\setminus;R_2$\\
 \hline
 Case I with light ${\bf \tl{H}_{3,8}}$&$\setminus;R_3$&$\setminus;R_3$&$(17.75,43.41)$&$(17.70,45.51)$\\
 \hline
 Case II with light ${\bf \tl{H}_{3,8}}$&$(17.98,30.03)$&$(16.32,36.88)$&$(15.52,40.44)$&$\setminus;R_2$\\
 \hline\hline
 \end{tabular}
\label{bm1}
\end{table}
%%%%%%%%%%%%%%%%%%%%%%%%%%%%%%%%%%%%%
\section{\label{sec-4} Proton decay triggered by flipped SU(6) GUT}
The instability of proton is one of the most striking consequence of the GUT models, which can possibly be tested by various experiments. Current lower limits on the lifetimes for many possible proton decay modes can be used to constrain various GUT models~\cite{proton:current,DUNE,JUNO:2015zny}. The nucleon decay via the exchange of GUT-breaking heavy gauge bosons is strongly suppressed by $M_{GUT}^4$ while the nucleon decay via the exchange of color-triplet Higgs is suppressed by $M^2_H$, which is the dominant proton decay mode for many SUSY GUT models even though it has additional suppression factors with light fermion Yukawa couplings and a loop factor. Here we concentrate on the non-SUSY case and leave the SUSY case in our subsequent studies. So, proton decay in our non-supersymmetric flipped $SU(6)$ model is mediated by dimension-6 operators, due to the exchanging of SU(6) gauge bosons.

 The relevant gauge interaction terms are given by
\beqa
{\cal L}&\supseteq& -\sqrt{2}g_6\[(\epsilon_{\alpha\beta} V_l(U_L^{c;A})^\da\ga^\mu X_{\mu; A}^\beta L_L^\al)+ \epsilon_{ABC} V_{CKM} (Q_L^A)^\da \ga^\mu X_{\mu;B} D_L^{c;C}\nn\right.\\
 &&~~~~~~~~~~\left.+ \epsilon_{\al\beta} V_N^\da (N_L^\al)^\da \ga^\mu X_{\mu;B}^\beta D_L^{c;B}  \],
\eeqa
with $X_{\mu; A}^\beta$ the heavy $SU(6)$ gauge bosons,  the index $'\beta=4,5'$ (corresponds to the $SU(2)_L$ index within $SU(3)_L$), $A=1,2,3$ the $SU(3)_c$ color indices and $V_{CKM},V_l,V_N$ being the mixing matrices for quarks, charged leptons and neutrinos, respectively.

Similar to flipped SU(5)~\cite{Ellis:2020qad}, below the flipped $SU(6)$ breaking scale, the dim-6 operator from integrating out the heavy gauge bosons of $SU(6)$ can be written as
\beqa
{\cal L}&\supseteq& \f{g_6^2}{M_X^2}\epsilon_{ijk}\epsilon_{\al\beta}\[C_1 (\bar{d}_{i L}^c \ga^\mu Q_{j\al L}) (\bar{u}_{k L}^c \ga_\mu L_{\bt L})+C_2 (\bar{d}_{i L}^c \ga^\mu Q_{j\al L})(\bar{N}_{L}^c\ga_\mu Q_{k \bt L})\],
\eeqa
with $C_1,C_2$ the relevant coefficients. As the RH neutrinos are much heavier than the proton, only the first operator will contribute to proton decay. The effective dim-6 operator that trigger the decay of protons takes the form
\beqa
{\cal L}(p\ra \pi^0~{l}_a^+)&\supseteq&
\f{g_6^2}{M_X^2}V_{CKM;11}^*(U_l)_{a1}\epsilon_{ijk}(u_R^i d_R^j) (l_{L a} u_L^k)~,
\label{pd:dim-6}
\eeqa
below the EW scale. The proton decay rate in our flipped SU(6) model can be calculated by generalizing the flipped SU(5) case~\cite{Ellis:2021vpp}
\beqa
\Ga(p\ra \pi^0 {l}_i^+)&=&\f{m_p}{32\pi}\(1-\f{m_\pi^2}{m_p^2}\)^2\left|{\cal A}(p\ra \pi^0 \bar{l}_i^+)\right|^2,\nn\\
&=&\f{g_6^4}{32\pi M_X^4}m_p|V_{ud}|^2|(U_{l})_{i1}|^2\(1-\f{m_\pi^2}{m_p^2}\)^2{\cal A}_L^2{\cal A}_{S1}^2\(\langle \pi^0|(ud)_R u_L|p \rangle_{l_i}\)^2,
\label{flip:proton}
\eeqa
with the two-loop perturbative QCD renormalization factor ${\cal A}_L = 1.247$.
The values ${\cal A}_{S1}$ of the renormalization factors for the dim-6 proton decay operator between the $M_X$ scale and the $m_Z$ scale are given with the following one-loop coefficients
\beqa
{\cal A}_{S1}&\approx&\[\f{\al_3(M_{331})}{\al_3(M_X)}\]^{\f{2}{5}}
\[\f{\al_3(M_{Z})}{\al_3(M_{331})}\]^{\f{2}{7}}
\[\f{\al_2(M_{331})}{\al_2(M_X)}\]^{\f{34}{27}}\[\f{\al_2(M_{Z})}{\al_2(M_{331})}\]^{\f{3}{4}}\nn\\
&&\[\f{\al_Y(M_{331})}{\al_Y(M_X)}\]^{-\f{33}{362}}\[\f{\al_Y(M_{Z})}{\al_Y(M_{331})}\]^{-\f{11}{84}}~,
\eeqa
for case I without light ${\bf \tl{H}_{3,8}}$ Higgs and
\beqa
{\cal A}_{S1}&\approx&\[\f{\al_3(M_{331})}{\al_3(M_X)}\]^{\f{6}{11}}
\[\f{\al_3(M_{Z})}{\al_3(M_{331})}\]^{\f{2}{7}}
\[\f{\al_2(M_{331})}{\al_2(M_X)}\]^{-\f{27}{2}}\[\f{\al_2(M_{Z})}{\al_2(M_{331})}\]^{\f{27}{28}}\nn\\
&&\[\f{\al_Y(M_{331})}{\al_Y(M_X)}\]^{-\f{33}{410}}\[\f{\al_Y(M_{Z})}{\al_Y(M_{331})}\]^{-\f{11}{96}}~,
\eeqa
for case II with light ${\bf \tl{H}_{3,8}}$ Higgs. Expressions of ${\cal A}_{S1}$ for other scenarios can be straightforwardly obtained.
The hadronic matrix elements can be obtained by lattice calculations~\cite{lattice}
\beqa
\langle \pi^0|(ud)_R u_L|p \rangle_{l_i}&=&\left\{\bea{c} -0.131,~~(l_1=e)\\-0.118, ~~(l_2=\mu)\eea\right.,
~\langle K^0|(u s)_R u_L|p \rangle_{l_i}=\left\{\bea{c}0.103,~~(l_1=e)\\0.099,~~(l_2=\mu)\eea\right.,\nn\\
\langle \pi^+|(u d)_R d_L|p \rangle&=&-0.186,~
\eeqa
within which the subscripts '$e$' and '$\mu$' indicate that the matrix elements are evaluated at the corresponding lepton kinematic points. Using eq.(\ref{flip:proton}), we can calculate the partial lifetime of the $p\to e^+ \pi^0$ mode in flipped SU(6) GUT models. The decay widths for other proton decay channels, such as $p\ra \pi^0 \mu^+$ and $p\ra K^0 e^+$ can be similarly obtained after taking into account proper CKM and PMNS matrix elements. The leptonic PMNS mixing matrix is given by~\cite{PMNS1,PMNS2}:
\begin{equation}
 V_{\rm PMNS} =
\begin{pmatrix}
 c_{12} c_{13} & s_{12} c_{13} & s_{13} e^{-i\delta} \\
 -s_{12} c_{23} -c_{12} s_{23} s_{13} e^{i\delta}
& c_{12} c_{23} -s_{12} s_{23} s_{13} e^{i\delta}
& s_{23} c_{13}\\
s_{12} s_{23} -c_{12} c_{23} s_{13} e^{i\delta}
& -c_{12} s_{23} -s_{12} c_{23} s_{13} e^{i\delta}
& c_{23} c_{13}
\end{pmatrix}
\begin{pmatrix}
 1 & 0 & 0\\
 0 & e^{i\frac{\alpha_{2}}{2}} & 0\\
 0 & 0 & e^{i\frac{\alpha_{3}}{2}}
\end{pmatrix}
~,
\end{equation}
where $c_{ij} \equiv \cos \theta_{ij}$ and $s_{ij} \equiv \sin
\theta_{ij}$ with the mixing angles $\theta_{ij} = [0, \pi/2]$, the
Dirac CP phase $\delta \in [0, 2\pi]$ and the Majorana phases $\al_2$ and $\al_3$ being set to vanish.
For normally ordered (NO) or inversely ordered (IO) neutrino hierarchical spectrum, the matrix elements of $U_l=V^*_{PMNS}V_\nu$ can be given as
\begin{align}
 (U_l)_{11}
&\simeq
\begin{cases}
 (V_{\rm PMNS}^*)_{11} = c_{12} c_{13} &\qquad \text{(NO)} \\[5pt]
 (V_{\rm PMNS}^*)_{13} = s_{13} e^{i\delta - i\frac{\alpha_3}{2}} &\qquad \text{(IO)}
\end{cases}
~, \label{eq:ul11}\\[3pt]
 (U_l)_{21}
&\simeq
\begin{cases}
 (V_{\rm PMNS}^*)_{21} =  -s_{12} c_{23} -c_{12} s_{23} s_{13}
 e^{-i\delta} &\qquad \text{(NO)} \\[5pt]
 (V_{\rm PMNS}^*)_{23} = s_{23} c_{12} e^{-i\frac{\alpha_3}{2}} &\qquad \text{(IO)}
\end{cases}
~. \label{eq:ul21}
\end{align}
We use the following best-fit value of the PMNS mixing and phase~\cite{NuFIT}
\beqa
\theta_{12}&=&33.82^{+0.78}_{-0.76},~\theta_{23}=48.3^{+1.2}_{-1.9},~\theta_{13}=8.61^{+0.13}_{-0.13},
~\delta=222^{+38}_{-28},~~~~~~(NO),\nn\\
\theta_{12}&=&33.82^{+0.78}_{-0.76},~\theta_{23}=48.6^{+1.1}_{-1.5},~\theta_{13}=8.65^{+0.13}_{-0.12},
~\delta=285^{+24}_{-26},~~~~~~(IO),
\eeqa
for IO and NO cases, respectively. Similar to ordinary flipped SU(5) case~\cite{2003.03285}, we can calculate the relations for partial decay widths
\beqa
\f{\Ga(p\ra \pi^0 \mu^+)}{\Ga(p\ra \pi^0 e^+)}
&\simeq &\f{\langle \pi^0|(ud)_R u_L|p \rangle_{\mu})^2}{(\langle \pi^0|(ud)_R u_L|p \rangle_{e})^2}\f{|(U_l)_{21}|^2}{|(U_l)_{11}|^2}
\approx \left\{\bea{c}0.114,~~~(NO)\\ 19.727,~~~(IO)\eea\right.~,\nn\\
\f{\Ga(p\ra K^0 e^+)}{\Ga(p\ra \pi^0 e^+)}
&\simeq &\f{\(\langle K^0|(u s)_R u_L|p \rangle_{e}\)^2}
{\(\langle \pi^0|(ud)_R u_L|p \rangle_{e}\)^2}\f{|V_{us}|^2}{|V_{ud}|^2}\approx 0.018~,\nn\\
\f{\Ga(p\ra K^0 \mu^+)}{\Ga(p\ra \pi^0 e^+)}
&\simeq &\f{\(\langle K^0|(u s)_R u_L|p \rangle_{\mu}\)^2}
{\(\langle \pi^0|(ud)_R u_L|p \rangle_{e}\)^2}\f{|V_{us}|^2|(U_l)_{21}|^2}{|V_{ud}|^2|(U_l)_{11}|^2}\approx \left\{\bea{c}0.004,~~~(NO)\\ 0.737,~~~(IO)\eea\right.~,\nn\\
\f{\Ga(p\ra \pi^+ \bar{\nu}_i)}{\Ga(p\ra \pi^0 e^+)}
&\simeq &\f{\(\langle \pi^+|(u d)_R d_L|p \rangle\)^2}
{\(\langle \pi^0|(ud)_R u_L|p \rangle_{e}\)^2}\f{1}{|V_{ud}|^2|(U_l)_{11}|^2}\approx \left\{\bea{c}3.151,~~~(NO)\\ 93.999,~~~(IO)\eea\right.~,
\label{PD:relations}
\eeqa
with $\Ga(p\ra \pi^+ \bar{\nu})=\sum\limits_{i}\Ga(p\ra \pi^+ \bar{\nu}_i)$. The $p\ra K^+\bar{\nu}_i$ channels can be proven to be forbidden by the unitary property of the CKM matrix. We also have
the relation ${\Ga(n\ra \pi^- l_i^+)}={\Ga(p\ra \pi^0 l_i^+)}/2$. By taking the ratio between
the two partial decay widths, many of the factors in the expressions (such as the $M_X$ scale, the SU(6) gauge coupling $g_6$ and the ${\cal A}_{L;S}$ factors) can be canceled, making the comparison of those ratios in various GUT models meaningful.

%%%%%%%%%%%%%%%%%%%%%%%%%%%%%%%%%%%%%
\begin{table}[htbp]\centering
 \caption{The partial proton decay lifetimes of the $(p\to e^+ \pi^0)$ mode in flipped SU(6) GUT for NO neutrino masses hierarchy. The partial proton decay lifetimes of this channel for IO case is related to that of NO case by $\tau_{IO}/\tau_{NO}\approx 30.1$. We calculate such partial life time for some benchmark points for various $M_{331}$ values in case I/case II with and without $M_{331}$ scale ${\bf \tl{H}_{3,8}}$ Higgs, respectively. The definitions of $'\setminus'$ symbol and the corresponding reasons $R_i$ are the same as that in  Table~\ref{bm1}.
 }
 \begin{tabular}{||c|c|c|c|c||}
 \hline\hline
 \small{$\tau(p\to e^+ \pi^0)_{\rm flipped}/yrs$} &\multicolumn{4}{|c|}{$M_{331}/{\rm GeV}$}\\
 \cline{2-5}
 $\tau_{NO}$&\small $3.16\tm 10^{3}$&$1.0\tm 10^{13}$ &\small $5.0\tm 10^{14}$&\small $3.16\tm 10^{16}$  \\
 \hline\hline
 \small Case I without ${\bf \tl{H}_{3,8}}$&\small $\setminus;R_3$&$\setminus;R_3$&\small $\setminus;R_3$&\small $6.92\tm 10^{46}$\\
 \hline
 \small Case II without ${\bf \tl{H}_{3,8}}$&\small $\setminus;R_1$&\small $1.04\tm 10^{45}$&\small $1.21\tm 10^{37}$&\small $\setminus;R_2$\\
 \hline
 \small Case I with ${\bf \tl{H}_{3,8}}$&\small $\setminus;R_3$&\small $3.88\tm 10^{43}$&\small $2.29\tm 10^{43}$&\small $1.61\tm 10^{43}$\\
 \hline
 \small Case II with ${\bf \tl{H}_{3,8}}$&\small $5.60\tm 10^{43}$&\small $8.21\tm 10^{35}$&\small $2.80\tm 10^{34}$&\small $\setminus;R_2$\\
 \hline\hline
 \end{tabular}
\label{proton:decay}
\end{table}
%%%%%%%%%%%%%%%%%%%%%%%%%%%%%%%%%%%%%\
\normalsize
We show in Tab.\ref{proton:decay} the numerical results of $p\to e^+ \pi^0$ partial life time for some benchmark points in case I/case II with and without $M_{331}$ scale ${\bf \tl{H}_{3,8}}$ Higgs for NO neutrino mass hierarchy case. The partial proton decay lifetimes of $p\to e^+ \pi^0$ for IO case is related to that of NO case by $\tau_{IO}/\tau_{NO}\approx 30.1$. We can see that without light ${\bf \tl{H}_{3,8}}$ Higgs, the partial proton decay lifetimes of the $p\to e^+ \pi^0$ mode are rather large. Although the current proton decay bound $\tau(p\to e^+ \pi^0) > 1.6\tm 10^{34}$ year~\cite{proton:current} can easily be satisfied, such proton decay mode cannot be observed in the forthcoming experiments. On the other hand, with $M_{331}$ scale ${\bf \tl{H}_{3,8}}$ Higgs, the partial proton decay lifetimes $p\to e^+ \pi^0$ can be as low as $2.80\tm 10^{34}$ years for $M_{331}\approx 5.0\tm 10^{14}$ GeV (of case II with NO neutrino hierarchy), although such a $M_{331}$ value lies near the upper bound on $M_{331}$.  It can be seen that such parameter regions for either cases  will soon be tested by future DUNE~\cite{DUNE} and Hyper-Kamiokande~\cite{Hyper-Kamiokande:2018ofw} experiments, which can reach as large as $1.3\tm 10^{35}$ years for $p\to e^+ \pi^0$ channel in future Hyper-Kamiokande.
%%Similarly, from Tab.\ref{proton:decay}, we can see that the benchmark parameter point that corresponds to $\tau(p\to e^+ \pi^0)=2.8\tm 10^{34}$ years for $M_{331}\approx 5.0 \tm 10^{14}$ GeV (of case II) can also be tested by the $p\ra \pi^0 e^+$ channel in future Hyper-K.

For other proton decay channels,
the only worrisome ones are $\tau(p\ra \pi^0 \mu^+)$ and $\tau(p\ra \pi^+ \bar{\nu}_i)$, with current $90\%$ CL lower bounds
\beqa
\tau(p\ra \pi^0 \mu^+)\gtrsim 7.7\tm 10^{33} {\rm yr},~~\tau(p\ra \pi^+ \bar{\nu}_i)\gtrsim 0.39\tm 10^{33} {\rm yr}~.
\eeqa
From the relations in (\ref{PD:relations}), we can see that the benchmark parameter point with $\tau_{IO}(p\to e^+ \pi^0)=8.4\tm 10^{35}$ years for $M_{331}\approx 5\tm 10^{14}$ GeV (of case II with IO neutrino hierarchy), corresponding to $2.8\tm 10^{34}$ years for NO neutrino hierarchy, can be ruled out by the future proton decay constraints for $p\ra \pi^0 \mu^+$ channel in Hyper-K, which can reach the sensitivities $6.9\tm 10^{34}{\rm yr}$ for this channel~\cite{Hyper-Kamiokande:2018ofw}. So, we can see that, for $M_{331}\sim 10^{15}$ GeV of case II, both NO or IO neutrino hierarchy predictions can be tested in the future proton decay experiments.

It is worthy to be noted that an additional effective dim-6 operator other than that given in~(\ref{pd:dim-6}) can contribute to the decay of proton in ordinary SU(6) GUT for $\emph{331}$ model. The Wilson coefficient for the dim-6 operator in~(\ref{pd:dim-6}) is also different to that of ordinary SU(6) GUT. Therefore, the predictions for the ratios of various proton decay modes in our flipped SU(6) model are different from that of ordinary SU(6) GUT for $\emph{331}$ model, leading to different predictions for both models.

\section{\label{sec-5}Conclusions}
We propose to partially unify the sequential $SU(3)_c\tm SU(3)_L\tm U(1)_X$ model (with $\beta=1/\sqrt{3}$) into a flipped SU(6) GUT model. It can checked that the gauge anomaly cancellation can be satisfied for each generation. We discuss the relevant Higgs sector, the low energy $\emph{331}$ model spectrum  and the unification of $SU(3)_c$ and $SU(3)_L$ gauge couplings. Neutrino masses generation and successful gauge coupling unification can set lower/upper bounds on the $\emph{331}$ breaking scale. The partial proton decay lifetime of various channels, for example, the $p\to e^+ \pi^0$ channel, in flipped SU(6) GUT are discussed.  We find that certain parameter region with $M_{331}\sim 10^{15}$ GeV of case II (for case with $M_{331}$ scale ${\bf \tl{H}_{3,8}}$ Higgs field) can predict a partial proton lifetime of order $10^{34}$ years for $p\to e^+ \pi^0$ mode, which can be tested soon by future DUNE and Hyper-Kamiokande experiments. The predictions for the ratios of various proton decay modes in our flipped SU(6) model are different from that of ordinary SU(6) GUT for $\emph{331}$ model.

It is known that there is a factor of approximately 20 to 25 between the unification scale for MSSM $M_{GUT}$ and the string scale $M_{str}$ in the weakly coupled heterotic string theory, which is given by $M_{str}\approx 5.27 \tm 10^{17} {\rm GeV}$ for the string coupling constant $g_{str}\sim {\cal O}(1)$. As string theory predicts "a prior" a unification of gauge couplings at $M_{str}$, such an inconsistency may indicates the appearance of (partial) unification at the intermediate scale $M_{GUT}$,  which would then unify with gravity and any other "hidden-sector" gauge symmetries at $M_{str}$~\cite{string:GUT}. In our model, for most benchmark points, the $SU(6)\tm U(1)_K$ breaking scale $M_X$ can be calculated to lie in the range ${\cal O}(10^{15}\sim 10^{18})$ {\rm GeV}. We should note that the calculated $M_X$ scale is sensitive to the additional Higgs contents (other than the minimal necessary $H_1(1,\bar{3},\f{2}{3})$ ,$H_2(1,\bar{3},-\f{1}{3})$,$H_3(1,\bar{3},-\f{1}{3})$ Higgs fields) adopted in the low energy $\emph{331}$ model. Taking into account the heavy string threshold corrections from the infinite towers of Planck-scale string states to RGE, the intermediate step $SU(6)\tm U(1)_K$ partial unification can be welcome to realize string scale unification of flipped SU(6) (or $SO(12)$/$E_6$ GUT group) with gravity.

We should note that non-sequential $\emph{331}$ model contains less exotic matter contents than the sequential $\emph{331}$ model. When the non-sequential $SU(3)_c\tm SU(3)_L\tm U(1)_X$ model is embedded into (partial) unification model, the flipped SU(6) gauge group can be advantageous over ordinary SU(6) gauge group in several aspects, for example, requiring much smaller representations for the embedding (thus much less exotic matter contents) and simpler anomaly cancellation constraints. The flipped SU(6) partial unification of non-sequential $\emph{331}$ model will be given in our future works~\cite{alternative:fsu6}.

\begin{acknowledgments}
We are grateful to the referees for helpful suggestions. This work was supported by the
Natural Science Foundation of China under grant numbers 12075213; by the Key Research Project of Henan Education Department for colleges and universities under grant number 21A140025.
\end{acknowledgments}
\appendix
\section{Appendix: Doublet-Splitting Problem in Flipped SU(6) Model}
\subsection{Missing partner mechanism in flipped $SU(6)$\label{appendix-1}}
Below the flipped $SU(6)$ breaking scale $M_X$, the colored Higgs fields can acquire masses of order $M_X$ while the uncolored ones can still be as light as $M_{331}$ scale. Such Doublet-Triplet (D-T) like splitting can be realized by the missing partner mechanism or by orbifold projection with proper boundary conditions.

 In the SUSY flipped SU(6) model, the missing partner mechanism can be applied by adopting the following type of superpotential
 \small
\beqa
W&\supseteq& \epsilon^{ijklmn}\la \({\bf 20_{\f{1}{2};H}}\)_{ijk}\( {\bf {15}^\pr_{0;H}}\)_{lm}\({\bf {6}_{-\f{1}{2};H}}\)_{n}+
\epsilon^{ijklmn}\la^\pr \({\bf \overline{20}_{-\f{1}{2};H}}\)_{ijk}\({\bf \overline{15}^\pr_{0;H}}\)_{lm}\({\bf \bar{6}_{\f{1}{2};H}}\)_{n}\nn\\
&+&M_{\bf 15^\pr}{\bf \overline{15}^\pr_{0;H} {15}^\pr_{0;H}}+X\({\bf 20_{\f{1}{2};H}}{\bf \overline{20}_{-\f{1}{2};H}}-M_X^2\)~,
\eeqa
\normalsize
for $M_{\bf 15^\pr}\sim M_X$.

 The F-flat and D-flat conditions can render the VEVs $\langle{\bf 20_{\f{1}{2};H}}\rangle=\langle {\bf \overline{20}_{-\f{1}{2};H}}\rangle=M_X$ (along the $(1, 1, \f{3}{2\sqrt{3}})_{\f{1}{2};H}$ directions) and break the flipped $SU(6)$ into $\emph{331}$ gauge group. The $(\bar{3},1,\f{1}{2\sqrt{3}})_{\f{1}{2};H}$ component within ${\bf \bar{6}_{\f{1}{2};H}}$ pairs to the $({3},1,\f{1}{\sqrt{3}})_{0;H^\pr}$ components within ${\bf \overline{15}^\pr_{0;H}}$ while the $ (1,\bar{3},-\f{1}{2\sqrt{3}})_{\f{1}{2};H}$ component cannot find any partner. Therefore, the colored component within ${\bf \bar{6}_{\f{1}{2};H}}$ is heavy while the uncolored one is much lighter ( at or below the $M_{331}$ scale). Missing partner mechanism of similar settings can also push heavy the colored components within ${\bf \bar{6}_{-\f{1}{2};H}}$ and ${\bf {15}_{0;H}}$.

\subsection{D-T splitting from orbifold projection\label{appendix-2}}

 Boundary conditions~\cite{orbifold1,orbifold2,hebecker} can also be used to split the colored and uncolored Higgs components. For example, we can adopt $S^1/Z_2$ orbifold by identifying the fifth coordinate $y$ under the two operations
\beqa
{\cal Z} : y \ra -y,~~~~{\cal T} : y \ra y + 2\pi R.
\eeqa
 Two inequivalent 3-branes located at $y=0$( $O$-brane) and $y=\pi R$ (${O}^\pr$-brane). We can choose the boundary condition with $P= diag(-1,-1,-1,~1,~1,~1)$ (under ${\cal Z}$ reflection) and $P^\pr=diag(~1,~1,~1,~1,~1,~1)$ (under the reflection ${\cal Z}^\pr= {\cal ZT}$) so as that the Higgs satisfy
 \beqa
 {\bf \overline{6}}_{\f{1}{2};H}&=&
(\bar{3},1,\f{1}{2\sqrt{3}})^{(-,+)}_{\f{1}{2};H}\oplus (1,\bar{3},-\f{1}{2\sqrt{3}})^{(+,+)}_{\f{1}{2};H},\nn\\
{\bf \overline{6}}_{-\f{1}{2};H}&=&
(\bar{3},1,\f{1}{2\sqrt{3}})^{(-,+)}_{-\f{1}{2};H}\oplus (1,\bar{3},-\f{1}{2\sqrt{3}})^{(+,+)}_{-\f{1}{2};H},\nn\\
{\bf 15}_{0;H}&=&(\bar{3},1,-\f{1}{\sqrt{3}})^{(-,+)}_{0;H}\oplus (3,3,0)^{(-,+)}_{0;H}
 \oplus (1,\bar{3},\f{1}{\sqrt{3}})^{(+,+)}_{0;H}~,\nn\\
{\bf 21}_{1;H}&=&({6},1,-\f{1}{\sqrt{3}})^{(-,+)}_{1;H}\oplus (3,3,0)^{(-,+)}_{1;H}
 \oplus (1,{6},\f{1}{\sqrt{3}})^{(+,+)}_{1;H}~,
 \eeqa
with proper brane localized terms to change some of the boundary conditions
from Neumann to Dirichlet.

\end{document}